\documentclass{article}
\usepackage{graphicx} 
\usepackage{braket}
\usepackage{amsmath}
\usepackage{amssymb}
\usepackage{geometry}
\usepackage{placeins}
\usepackage[subtle]{savetrees}
\author{Robert A. Lawrence $^1$}

\date{$^1$ School of Physics, Engineering and Techonology, Univerist yof York, Heslington, Y010 5DD \\ \today}
\title{Are Magnons just the van der Waals interaction in Disguise?}
\begin{document}

\maketitle
\abstract{The MBD model of the van der Waals interaction is extended to also consider magnetic interactions, and it is demonstrated how this can be made to reproduce the Heisenberg Hamiltonian. It is found that this leads to a weak coupling between the charge dipole waves that are the basis for the electric-only van der Waals interaction and the spin-dipole waves (magnons) of the Heisenberg model. By applying the same level of theory to both simultaneously we demonstrate that magnons and  charge-dipole waves may both be considered to be a basis for the fluctuating part of the many-body electron density. }

\section{Introduction}

One of the enduring appeals of magnetic materials is their capacity for use in high energy efficiency devices; this is the main driver behind the recent surge of interest in materials for spintronics and magnonics. Devices made using this framework promise higher performance for lower energy cost --  both improving consumer experience and helping to combat climate change through improved energy efficiency.

Ultimately, the development of these devices will require the discovery of new materials: both to enable miniaturisation (some materials require thick layers to demonstrate desirable properties \cite{VallejoFernandez2021}) and also to make the process more economical: much of the current generation of spintronic technology relies on alloys of Ir and Mn, the former of which is incredibly rare within the Earth's crust, and accordingly expensive:  thereby limiting the expansion of spintronics technology until Ir can be replaced.

An efficient search for new materials with desirable properties is vastly enhanced by even a basic theoretical understanding of the origins of those properties. This enables the selection of more viable candidate materials for testing thereby saving much wasted research efforts into potential candidate materials which a simple theoretical consideration might eliminate early on.

Building up these insights generally requires being able to correlate both structural and chemical features of the candidates with the theoretical values of their properties. Such a programme of research would currently require significant computational expense: the two major methods for evaluating magnetic properties of systems are to either use a many-body (MBPT) or time dependent-density functional perturbation theory (TD-DFPT) approach \cite{TurboMagnon} or else to perform many finite displacements (FD) within a density functional theory (DFT) approach. All of these methods suffer from having a very high computational cost attached. 

It is also worth noting that that the micromagnetics and atomistic spin dynamics communities -- which are both more well-established than the \emph{ab initio} magnetic community and much more numerous -- tend to rely on the Heisenberg Hamiltonian to describe the magnetic properties of the materials they study. This has the advantages of being very simple and provides a good intuitive basis on which to consider magnetic interactions. Equally, it is a phenomenologial model replete with material-dependent parameters, which makes it next-to-useless for materials discovery purposes: if you don't know the parameters for the material then you have no information at all! This in turn makes evaluating the material-dependent constants an important application of \emph{ab initio} theory within the magnetics community.

\subsection{Magnons as a basis set}

One perspective on magnons (or spin waves -- the author uses magnons to refer to both pictures), is as a basis set to describe the dynamical spin excitations of a system. Once this basis set is known, temperature dependence may be simply calculated through the use of statistical mechanics, and higher order terms may be represented via magnon-magnon coupling as a combination of this underlying basis. 

Magnons are \emph{spin} dipole waves typically described using the Heisenberg Hamiltonian \cite{Barker2020,Lack2024,Szilva2023}:  

\begin{equation}\label{eqn:Heis_Ham}
    \hat{H} = \sum_i \hat{S}_i \cdot \textbf{K}_{i} \cdot \hat{S}_i + \sum_{i\neq j} \hat{S}_i \cdot \textbf{J}_{ij} \cdot \hat{S}_j
\end{equation}
where we have compressed Heisenberg exchange, two-site anisotropy, the Dyalozhinskiy-Moriya interation (DMI) and the variation of  spin magnitude with orientation into  the tensor quantity $\textbf{J}_{ij}$ \cite{Szilva2023}, and have chosen to express the anisotropy in a general representation as the tensor $\textbf{K}_{i}$, to represent the anisotropy of moving a single spin without needing to pre-define the functional form (uniaxial, cubic, hexagonal, etc) of the anisotropy. 

The current large challenge, however, for first principles magnetism is properly describing these tensor parameters at all: these effects are typically both small (requiring a high accuracy calculation) and long range (requiring a large supercell in plane-wave DFT -- the workhorse of the DFT world). Together these frequently make calculations prohibitively expensive and also limit them to bulk systems. This is clearly a drawback as the most important regions to investigate are not bulk-like, but rather near heterointerfaces, such as might be found in a real magnetic device.

In this paper, we will explore a way of quantifying the magnon parameters through the use of an auxiliary model, in the same spirit as has already been done successfully many times for the van der Waals (vdW)  interaction \cite{Kim2015,Carrasco2013,Amft2011,Yuan2020}. 

\subsection{Static and Dynamic Correlations}

The many-body Schr\"odinger (or Dirac) equation is notoriously hard to solve; and indeed impossible to solve analytically for a system with a large number of electrons. Therefore cheaper \emph{ab initio} methods such as Density Functional Theory (DFT) have been developed in order to solve this problem. Whilst the DFT framework is formally exact, in practice the universal exchange-correlation (XC) functional is unknown, and therefore approximations must be made to capture the physics of this universal functional. This makes the quality of a DFT simulation critically dependent on choosing an appropriate functional for the system that captures all of the relevant physics.

Conversely, the single most accurate model of the many-electron wavefunction is the configuration-interaction (CI) model in the limit of a complete basis set ("Full CI") \cite{Sherrill1998}. Whilst in practice this method is prohibitively costly for all but the simplest molecules, it is still instructive to think about to highlight the origin of some of the effects relevant to magnons.

In the CI picture, the wavefunction may be written as a sum over the minimum energy unexcited state, all singly-excited states, all doubly-excited states and so on. This wavefunction may be expressed as

\begin{equation}\label{eqn:CI}
    \ket{\Psi} = c^0\ket{\psi^0} + \sum_i c_i^1 \ket{\psi_i^1} +\sum c_i^2 \ket{\psi_i^2} + \dots 
\end{equation}
where $\ket{\Psi}$ is the total many-body wavefunction, $\ket{\psi^k_i}$s are individual Slater determinants with $k$ indicating the number of excited electrons, and $i$ indicating which particular combination of excitations is involved, finally $c_i^k$ are coefficients controlling the contribution to the total wavefunction of that individual $\psi_i^k$.

The key point, however, is that this is a \emph{mixed} state, meaning that the ground state of this system evolves over time. However, unlike many wavefunctions, it is typically found that $c^0$ dominates over the rest of the coefficients, such that the wavefunction may be effectively partitioned into a quasi-static part and a quasi-dynamic part. Within density functional theory, the effects of the higher order Slater determinants that contribute to the static components of the wavefunction are bought in as part of the correlation functional (marking the key difference between DFT and Hartree-Fock theory), whereas the dynamical contributions are typically neglected. 

By way of analogy, first heard by the author in an excellent talk by A. Tkatchenko\cite{psi_k_TK_talk} one might consider this to be like a swarm of flies around an ox: na\"ively the impression from afar is just the presence of the ox (the static contribution to the density), however on closer inspection one can see the whole system also consists of a swarm of flies, forever moving in a correlated way with each other and the ox itself, and furthermore that the ox might react (by swishing its tail!) to the presence of the swarm of flies. 

For many systems, the approach of only considering the static density is sufficient to get an accurate description of the electronic structure and the related properties of interest -- for example bulk moduli -- and the dynamic contributions may safely be discarded without affecting the validity of the physical model. For other systems (including molecular systems and van der Waals heterostructures where non-bonding interactions matter) these dynamical effects are critical to the last few percent of accuracy which can qualitatively affect the answer, for example the inter-layer spacing in graphite.

To evaluate these properties correctly, the extra dynamical correlations must be added in to describe the correct physics (typically using semi-empirical dispersion corrections, although explicit vdW XC-functionals are also used \cite{Graziano2012}). One consequence of these models is the spontaneous emergence of charge-dipole waves throughout the material \cite{Tkatchenko2014,Ambrosetti2016,Gori2023}: a quantised basis for the vdW (charge dipole) excitations of the density.

These semi-empirical dispersion models typically consist of an auxiliary system, frequently a set of coupled fluctuating dipole moments \cite{Tkatchenko2012} that coarse-grain the density into regions that may be represented by a single oscillating dipole moment. These dipole moments are then coupled together using classical electrodynamics, at which point an accurate parameterisation of the oscillators -- typically by linking to a beyond-DFT theory which may capture dynamical correlations -- is the only remaining challenge.

These models offer significant improvements in the treatment of the van der Waals interaction bringing properties such as the interlayer spacing of graphite or energies of adsorption of a molecule onto a substrate into quantitative agreement with experiment \cite{Graziano2012} as opposed to qualitative disagreement.

\section{The MBD model of the Van der Waals Interaction}
\FloatBarrier
Whilst many different successful models exist to model the van der Waals interaction, we choose to highlight here the Many-Body Dispersion (MBD) method of Tkatchenko \emph{et al.}  \cite{Tkatchenko2012} as being relatively transparent as well as both highly actively developed \cite{Tkatchenko2012,Tkatchenko2014,Bryenton2023} and accurate \cite{Ambrosetti2014}.

Many excellent review papers exist \cite{Grimme2016,Kim2015} that describe both MBD and other models in great depth, but in short the process is as follows:

\begin{enumerate}
    \item Divide the system into a set of localised fragments
    \begin{itemize}
        \item[$\rightarrow$] Typically these are atom-centered, although there is no obligation for this choice
    \end{itemize}
    \item For each fragment, calculate the effective polarisability based on scaling arguments for polarisability vs fragment volume and appropriately adjusting beyond-DFT values of the free atom's polarisability
    \item Perform a self-consistent screening (SCS) cycle to account for the anisotropy due to the local environment
    \item Use classical electrodynamics to model the interaction between different oscillators within a coupled fluctuating dipole moment model
    \item Diagonalise the model's Hamiltonian to extract the eigenmodes of the system (the charge dipole waves) and their eigenenergies
    \item Utilise this model Hamiltonian alongside the main DFT calculation to introduce the effects of the vdW correlations.
\end{enumerate}

\begin{figure}
    \centering
    \includegraphics[width=0.3\columnwidth]{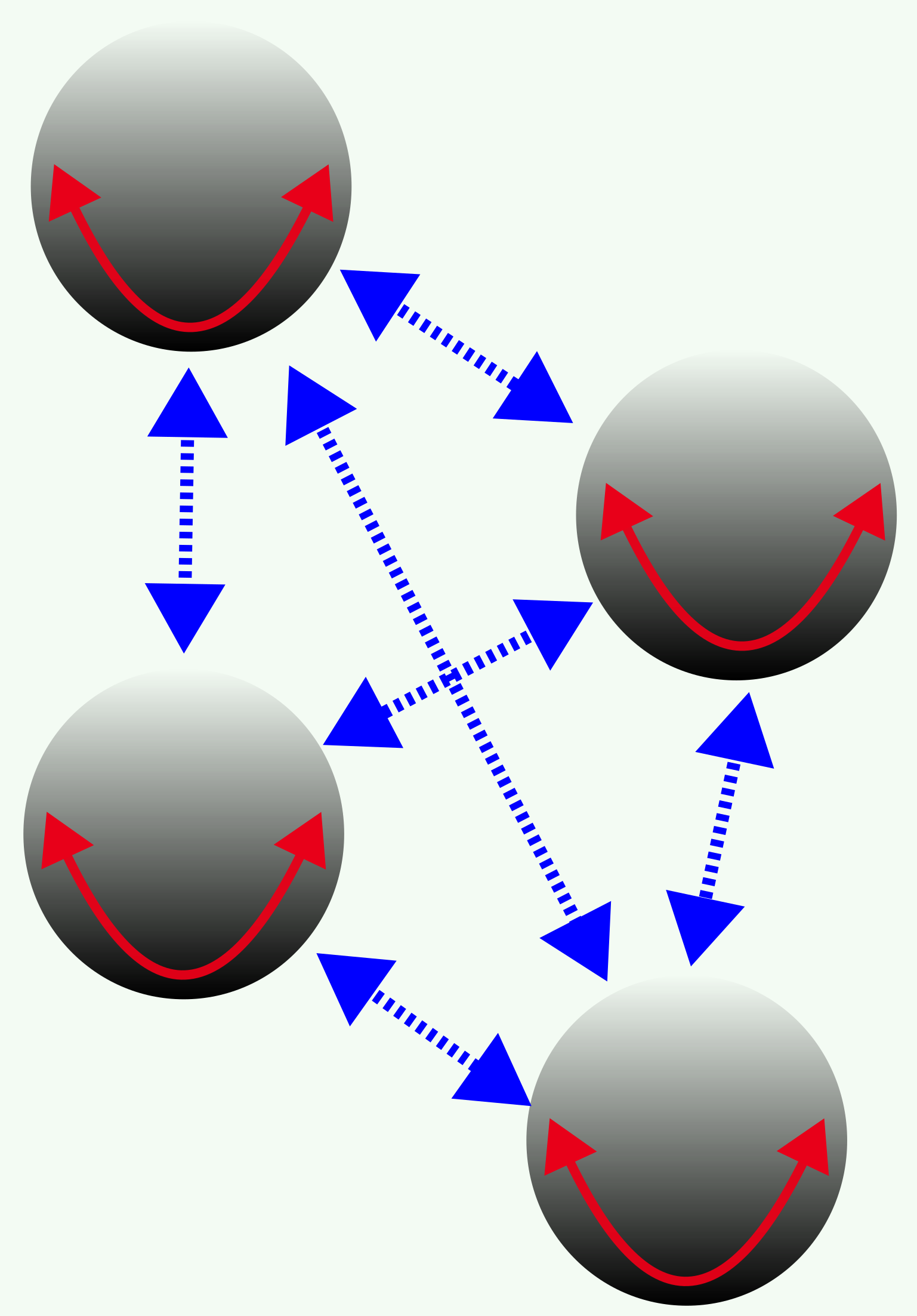}
    \caption{A schematic of the MBD model: individual  oscillators interact in a pairwise manner with all other oscillators in the system, leading to indirect interactions, or screening. This also causes the lowest energy excitations of the system to be dispersed waves rather than local oscillations.}
    \label{fig:MBD-coupling fig}
\end{figure}

This model is represented pictorially in figure \ref{fig:MBD-coupling fig}.

\FloatBarrier
The final MBD Hamiltonian may be written as \cite{Ambrosetti2016}

\begin{equation}\label{eqn:MBD_Ham}
    \hat{H}_{MBD} = -\frac{1}{2} \sum_{i=1}^{N} \nabla^2_{\vec{d}_i} +\sum_{i=1}^{N} \omega^2_i \vec{d}^2_i + \sum^N_{i<j}\omega_i\omega_j \sqrt{\alpha^0_i\alpha^0_j}\vec{d}_i T_{ij} \vec{d}_j
\end{equation}
where $d_i =\sqrt{m_i}\cdot p_i$ ($p_i$ is the dipole moment on site i and $m_i$ is the effective mass of the oscillator), $\nabla^2$ is the kinetic energy operator, $\omega_i$ is a characteristic frequency of the oscillator related to its force constant ($\omega = \sqrt{\frac{k_i}{m_i}}$ for force constant $k_i$ and mass $m_i$), $\alpha^0_i$ is the static contribution to the electric dipolarisability of the system and $T_{ij}$ is the dipole tensor to describe the interaction between the dipoles on sites i and j respectively. In principle higher order terms, such as the Axel-Teller-Muto (3-body) term exist, but the MBD model truncates at pairwise interactions.

This may be thought of as a sum over the kinetic energy, the one-body potential energy of the system (how much energy is required to rotate a single dipole moment) and the interaction between neighbouring dipole moments (as a function of their relative magnitude, orientation and separation).

\subsection{Self Consistent Screening}

It should be noted that in the isotropic limit, the second term of the MBD Hamiltonian, the one body potential, only affects the energy of the system when the magnitude of the dipole moment changes. This is not purely the case, however: the potential has no obligation to be isotropic, and this in turn leads to a coupling between the orientation of the dipole and its magnitude, introducing a directional dependence as well. Now, since the functional form of the potential remains the same, this anisotropy must originate in the effective force constant being different for the different orientations of the dipole moment.

The anisotropy of this force constant has two potential origins: firstly, the intrinsic polarisability of the fragment can have some anisotropy due to its underlying structure: which will be influenced by the electrostatic (and magnetic) fields of the static component of the density, since within our model each fragment is parameterised from a model which already includes all of these static correlations and any charge redistribution within the system. This term has typically been found to make only a small contribution to the anisotropy of many systems, since the perturbation of the atomic spherical potential by the electric fields of neighbouring atoms is typically weak.

The second origin is in the difference between the polarisability of the isolated fragment and the fragment-in-system. This is an emergent feature of the many-body interaction due to the screening of the polarisation by other oscillators. One way of thinking about this is as the impact of delocalised bands/orbitals which may contribute at multiple sites and therefore couple those multiple sites together: which has to be included somehow within the coupled oscillator model. This is done through the self-consistent screening (SCS) method. Within the MBD$_{@SCS}$ method, this second term is responsible for the dominant contributions to the anisotropy\cite{Tkatchenko2012}, and may also be expected to dominate the contributions to Heisenberg K. 

It should be noted, however, that the SCS method does have some drawbacks: notably the strong dependence of the final answer on the initial approximation for the non-interacting fragment's polarisability \cite{Bryenton2023}, which is currently an unresolved issue within the van der Waals community; nevertheless, it is agreed that the SCS process does reproduce the expected physics better than neglecting it \cite{Bryenton2023}.

In the SCS model, a Gaussian polarisation density is placed on each of the oscillators and these are allowed to interact through classical electrodynamics, giving \cite{Buko2013,Bryenton2023}

\begin{equation}
    \alpha_i^{\text{SCS}}(i\omega) = \alpha_i^0(i\omega) - \alpha_i^0(i\omega) \sum_{i\neq j}\textbf{T}_{ij} \alpha_j^{\text{SCS}}
\end{equation}
where $\alpha_i^{\text{SCS}}$ is a tensor object containing the self-consistently screened polarisability and the dipole-dipole interaction tensor $\textbf{T}_{ij}$ is given by \cite{Bryenton2023}

\begin{equation}
    \textbf{T}_{ij} = \nabla_{R_i} \otimes \nabla_{R_j} \nu_{ij}
\end{equation}
with the interaction potential $\nu_{ij}$ given by 

\begin{equation}
    \nu_{ij} = \frac{\text{erf}\frac{R_{ij}}{\sigma_{ij}}}{R_{ij}}
\end{equation}
where $\sigma_{ij}$ is the Pythagorean combination of the two $\sigma_q$ for $q \in \{i,j\}$ where $\sigma_q = (\sqrt{\frac{2}{9\pi}} \alpha_q^0)^\frac{1}{3}$ is the Gaussian distribution of the charge / spin for each of the harmonic oscillators.

\subsection{Double Counting and range-separation}

So far, the auxiliary model does not depend on the context in which it is to be implemented. The MBD model, however, is intended to be used to add extra information into DFT calculations, therefore it is also worth noting that one needs to take a degree of care when adding an auxiliary model to a DFT simulation to ensure that there is no double counting of the correlation. As one might expect, the short-range correlations (as included within DFT) already include some of the dynamic correlations being added by the auxiliary Hamiltonian.

To avoid this double-counting, the range separated self-consistent screening (RSSCS) method was developed, where a smooth function is used to  reduce and remove the effects of the auxiliary Hamiltonian within a van der Waals' radius of the centre of the ionic fragment. It should also be noted that at short ranges higher order multipole effects start to become relevant. The TS@SCS-MBD model \cite{Tkatchenko2012} neglects these higher order terms, but other models of  the van der Waals interaction such as the exchange dipole moment model \cite{Becke2006} explicitly include them. 

Notwithstanding this, when extending the MBD model to also include magnetism, we neglect this range-separation in the interests of simplicity and drawing qualitative conclusions from a general Hamiltonian (that does not require interfacing with DFT). This will lead to a decrease in the accuracy of the short-range components of the model -- in particular the nearest neighbour J couplings may be of poor quality without this corrective term. Introducing range separation within the electromagnetic vdW model is a matter for future work. 

\subsection{The Emergence of Charge Dipole Waves}

Now that the model is fully parameterised, we can see that the structure of the Hamiltonian for a three-body problem may be represented as
\begin{equation}
    \hat{H} =  \vec{D} \cdot \begin{pmatrix}
        V_{11} & V_{12} & V_{13} \\
        V_{12} & V_{22} & V_{23} \\
        V_{13} & V_{23} & V_{33}
    \end{pmatrix} \cdot \vec{D}
\end{equation}
where $\vec{D}$ is a vector consisting of the polarisation on each ionic centre and the $V_{ij}$ terms are the potential for the interaction between sites $i$ and $j$ (where $i=j$ is allowed) as parameterised in the MBD Hamiltonian. 

Now, as for any Hamiltonian, if we wish to determine the eigenstates of the system (which form the basis of any excited state of the model), we simply have to diagonalise it. This will yield a bosonic basis set, as is the case for the canonical quantisation of any such theory. Now, so long as the matrix isn't already diagonal (i.e. the two-body terms are non-zero), we find that the eigenbasis for the excitations of this system are delocalised over many atoms in the form of charge-dipole waves. 

The extra correlation energy associated with the van der Waals interaction is then given by the difference in the energies between the zero point energies of the coupled and uncoupled modes \cite{Tkatchenko2012}

\begin{equation}
    E_{vdW} = \frac{1}{2} \sum_{i=1}^3N \sqrt{\lambda_i} - \frac{3}{2} \sum_{i=1}^N \omega_i^{SCS}
\end{equation}

where $\lambda_i$ are the eigenvalues of the coupled Hamiltonian and $omega_i^{SCS}$ are the  zero point energies of the uncoupled oscillators. Within density functional simulations, this property and its derivatives may be used to apply corrections to both the total energy and evaluate the forces induced by the vdW interaction.

\section{Extending Van der Waals Models to magnetic systems}

For a simple carbon-based system (the most common use case for vdW corrections), one does not anticipate that magnetic effects will play a significant role: magnetic moments are typically small in these molecules, and the coupling between electric and magnetic degrees of freedom are minimal due to the tiny spin-orbit coupling (SOC) present in this class of systems.

Nevertheless, it is reasonable to expect a similar interaction to occur for magnetic moments interacting through magnetic dipoles, and by inspection of the Hamiltonians (eqns \ref{eqn:Heis_Ham} and \ref{eqn:MBD_Ham}) a similarity in the structure may be seen, with the exception of the kinetic energy term in the MBD Hamiltonian (eqn \ref{eqn:MBD_Ham}) which is not currently reproduced within the Heisenberg model. 

One consequence of this lack of kinetic energy treatment is that the final result of the Heisenberg Hamiltonian is a localised moment description with no inclusion of the so-called \emph{itinerant magnetism} \cite{Moriya1984}. Taken in isolation, the kinetic energy term will lead to a free electron like model, leading to delocalised contributions to the Hamiltonian, thereby including both itinerant and local magnetism on the same footing.

The remaining terms of the Hamiltonians (which are common to both) are an on-site (one-body) potential for the oscillators, and a pairwise inter-site coupling representing the interaction of the fields from one dipole with the dipole at the other site. 

To some extent the similarity between these two models is not surprising: both Hamiltonians are describing long-range correlations in the dynamical part of the electron (charge or spin) density through the use of coupled fluctuating dipole moment models. Then add in the fact that electricity and magnetism are two different aspects of the same \emph{electromagnetic} force, and it suddenly seems very reasonable that a unification between these two models is not only possible but desirable.

The key assumption here is that the expected properties of the fragment of the wavefunction being represented by the harmonic oscillator are uniquely defined by that wavefunction. This has the practical upshot that if one wishes to change the expectation of one property of the fragment (either by applying an external field, or as a local spontaneous excitation), then the wavefunction changes, which \emph{may} also lead to other properties of that fragment changing. 

\subsection{The Role of Spin-Orbit Coupling}

Even valence electrons can experience a degree of spin-orbit coupling in the presence of heavy atoms, because  their radial wavefunctions have a small but finite contribution at small radii (near to the nucleus). Importantly, this means they gain a significant amount of kinetic energy in this region and therefore are more accurately modelled with the non-relativistic limit of the Dirac equation (the Pauli equation) rather than the Schr\"odinger equation. This has the additional effects of converting the wavefunction at any one point from a complex scalar to a complex spinor object and introducing spin-orbit coupling which links the spin and charge degrees of freedom (this coupling is essentially just a question of reference frames at near-relativistic momenta). 

Accordingly, all of the work to solve this model is envisaged as being carried out in a non-collinear magnetic space (i.e. dealing ith spinor wavefunctions and a density which may meaningfully be decomposed onto all four -- including identity -- of the Pauli matrices) with the effects of spin-orbit coupling included. If this is not the case, then it is not possible to evaluate the coupling parameters between spin and charge degrees of freedom.

\subsection{Calculating the Generalised Static Polarisability}

Now, taking the MBD model as our starting point (since all of its parameters are well-defined in terms of fundamental constants (if indirectly!) rather than in terms of generalised Heisenberg K's and Heisenberg J's for a given material, we may attempt to unify the model by replacing every purely electric interaction with a fully electromagnetic one, thereby including magnetism within the MBD model of the van der Waals interaction.

Firstly, we define a pseudovector equation

\begin{equation} \label{eqn:gen_pol}
   \vec{P} = \begin{pmatrix}
        d_x \\[0.5em] d_y\\[0.5em] d_z \\[0.5em] \frac{S_x}{c} \\[0.5em] \frac{S_y}{c} \\[0.5em] \frac{S_z}{c}
    \end{pmatrix}
    = \begin{pmatrix}
       \mathbf{\alpha}_{ee} & \mathbf{\alpha}_{em}\\
       \mathbf{\alpha}_{me} & \mathbf{\alpha}_{mm} 
    \end{pmatrix}
    \begin{pmatrix}
        \frac{E_x}{c}\\[0.5em] \frac{E_y}{c}\\[0.5em] \frac{E_z}{c} \\[0.5em] B_x \\[0.5em] B_y \\[0.5em] B_z
    \end{pmatrix}
\end{equation}
where $\vec{P}$ is the 6-D electromagnetic polarisation, which is used to indicate the \emph{change} from the ground state. It is worth emphasising that any underlying permanent moments (i.e. the ground state spin structure) is not included within this vector and may be considered as an arbitrary offset to the central positions of the harmonic oscillators within this model.

The 6-D induced polarisation-magnetisation vector, $\vec{P}$ is related to the pseudovector for the applied electromagnetic field by a quantity we term ``generalised polarisability'', $\mathbf{\alpha} =\begin{pmatrix}
       \mathbf{\alpha}_{ee} & \mathbf{\alpha}_{em}\\
       \mathbf{\alpha}_{me} & \mathbf{\alpha}_{mm} 
    \end{pmatrix}$, which has components that may be expressed as
\begin{align}
    \mathbf{\alpha}_{ee} = \begin{pmatrix}
        \alpha_{xx} & \alpha_{xy} & \alpha_{xz} \\ 
        \alpha_{yx} & \alpha_{yy} & \alpha_{yz} \\ 
        \alpha_{yx} & \alpha_{zy} & \alpha_{zz} \\ 
    \end{pmatrix} \, , \,   \mathbf{\alpha}_{em} = \begin{pmatrix}
        \beta_{xx} & \beta_{xy} & \beta_{xz} \\ 
        \beta_{yx} & \beta_{yy} & \beta_{yz} \\ 
        \beta_{yx} & \beta_{zy} & \beta_{zz} \\ 
    \end{pmatrix}
\end{align}
and 
\begin{align}
\mathbf{\alpha}_{me} = \begin{pmatrix}
        \gamma_{xx} & \gamma_{xy} & \gamma_{xz} \\ 
        \gamma_{yx} & \gamma_{yy} & \gamma_{yz} \\ 
        \gamma_{yx} & \gamma_{zy} & \gamma_{zz} \\ 
    \end{pmatrix} \, , \, \mathbf{\alpha}_{mm} = \begin{pmatrix}
        \chi_{xx} & \chi_{xy} & \chi_{xz} \\ 
        \chi_{yx} & \chi_{yy} & \chi_{yz} \\ 
        \chi_{yx} & \chi_{zy} & \chi_{zz} \\ 
    \end{pmatrix}
\end{align}
where $\alpha_{ij}$ are components of the polarisability tensor, $\chi_{ij}$ are components of the magnetic susceptibility tensor, and $\beta$ and $\gamma$ are tensors to represent the induced electric polarisation caused by an applied magnetic field and the induced magnetisation caused by an applied electric field respectively. We also will use the notation $\vec{F} =    \begin{pmatrix}
        \frac{E_x}{c}& \frac{E_y}{c}& \frac{E_z}{c} & B_x & B_y & B_z
    \end{pmatrix}^\text{T} $ as a shorthand for the electromagnetic pseudovector. For a system with no spin-orbit coupling (as is approximately true for atoms with a low atomic number), one expects all components of $\beta$ and $\gamma$ to be 0, whereas for a system with strong SOC they become small but finite to represent the coupling between the charge and spin densities. 

It is worth emphasising that the $\beta$ and $\gamma$ tensors indicate the response of the electric components of a system to an applied electric field and the magnetic components of the system to an applied electric field respectively, which is only possible in the presence of spin-orbit coupling (SOC). This is because in the classical limit (no SOC) there is no link between the spin and charge degrees of freedom, whereas in the relativistic limit these two degrees of freedom are linked through the SOC.

Within the TS scheme, the ansatz that the electric polarisability is correlated strongly with the effective volume of the atom is used. This may be justified by considering that the greater the number of electrons present the greater the corresponding atomic volume and polarisability.

For magnetic materials, however, the situation is much more complicated: adding or removing an electron can both increase \emph{and} decrease the magnetisability since depending on the original electronic structure both of these can increase or decrease the magnetic moment of the local ion. This means that magnetic materials do not have a neat relationship between magnetisability and atomic volume, so an equivalent of the TS scheme cannot be used for magnetisability.

It is desirable to use one single method to evaluate both of the electric and magnetic contributions to the generalised polarisability in order to ensure a self-consistent model, and this method must involve the electronic structure of the chosen fragment in order to evaluate the magnetisability. 

Accordingly, we suggest that a perturbation theory approach -- whilst significantly more expensive than the simple model used in the Grimme \cite{Grimme2016,Caldeweyher2020} and TS \cite{Tkatchenko2009} approaches -- may have sufficient accuracy to also capture the magnetic properties of the system.

Accordingly, working from the definition that $\alpha = \frac{d\vec{P}}{d\vec{F}}$, we suggest that by calculating the first order wavefunction response to an infinitesimal applied field along each of the components of $\vec{F}$ it is possible to construct 6 new wavefunctions, $\psi^1_{F_i}$ from standard perturbation theory (with appropriate attention to any degeneracies) 

\begin{equation}
    \psi^1_{F_i} = \sum_{m\neq n} \frac{\bra{\psi_n^0}\hat{H}(F_i)\ket{\psi_m^0}}{E_m - E_n}\ket{\psi_n^0}
\end{equation}
where $\psi^1_{F_i}$ is the first order correction to the wavefunction due to the application of perturbing Hamiltonian $\hat{H}(F_i)$ associated with each component of $\vec{F}$. For the electric components of the field, this perturbing Hamiltonian is the scalar potential, $\Phi$, associated with an infinitesimal applied field, whereas for the magnetic components of the field, the appropriate potential is the vector potential, $\vec{A}$, which is introduced by transforming the momentum operator, $\vec{p} = -i\hbar\vec{\nabla} \rightarrow -i\hbar\vec{\nabla} + e\vec{A}$ where $e$ is the electronic charge and by adding in a Zeeman-like potential term, $\frac{e\hbar}{2m_e}\vec{s}\cdot \vec{B}$.

We note that evaluating the response of the XC-functional term to the perturbation is a technical challenge, which may benefit from further work, however existing schemes to perform this do exist \cite{Ricci2019}. 
We note that by applying infinitesimal fields, as is correct for a perturbative appropriate to evaluate the generalised polarisability, the problem with the large quantum of field in small periodic cells is avoided.

Now, noting that the change in expectation value for any property, $\langle O \rangle$ that can be evaluated by an operator, $\hat{O}$ of a system that has been perturbed, may be expressed as $\bra{\psi^1}\hat{O}\ket{\psi^1}$ (in short,  the change of expectation value is the expectation value of the change of the wavefunction). Now, by performing a projection onto a localised basis set before evaluating the property -- such as maximally localised Wannier functions \cite{Marzari2012} or atomic orbitals -- we may calculate the change of a fragment centered at the atomic center (or Wannier function centers if these do not align with the ionic positions) due to the infintesimal applied field; these will yield each of the individual components of the generalised polarisabilty of the fragment in question.

\begin{align}
    \mathbf{\alpha}_{ee, ij} &= \bra{\phi^1_{E_j}}  \vec{r}_i \ket{\phi^1_{E_j}}\\
    \mathbf{\alpha}_{em, ij} &= \bra{\phi^1_{B_j}} \vec{r}_i \ket{\phi^1_{B_j}}\\
    \mathbf{\alpha}_{me, ij} &= \bra{\phi^1_{E_j}} \vec{S}_i \ket{\phi^1_{E_j}}\\
    \mathbf{\alpha}_{mm, ij} &= \bra{\phi^1_{B_j}} \vec{S}_i \ket{\phi^1_{B_j}}
\end{align}
where $\ket{\phi^1_{F_q}}$ is the local projection of the first order correction to the wavefunction due to the $q^{th}$ component of the applied electromagnetic field, $\vec{r_i}$ is the component of the position operator in direction $i$, and $\vec{S}_i$ is the Pauli matrix in direction $i$.   

One advantage of only projecting $\ket{\psi^1}$  onto a local basis $\ket{\phi^1}$  following the system-wide perturbation is that the expensive part of the calculation (performing the perturbation theory) need only be performed 6 times -- once for each component of $\vec{F}$ -- and can capture non-local effects such as charge transfer between ionic sites. At the same time, by evaluating the change of polarisation on localised fragments, which have a well-defined origin, we avoid the cost of evaluating the change in polarisation using the modern theory of polarisation \cite{Spaldin2012} and can instead simply integrate the appropriate operator over the first order correction to the spin or charge density to find the change in property.

The author notes that this is a linear-response treatment of the problem, and that a more accurate model may, in principle, be obtained by calculating the non-linear responses in addition to the purely linear responses considered here, although this is much more computationally demanding task \cite{Alliati2023}. This would simply involve extending the model from a set of harmonic oscillators to anharmonic ones instead.

We also note that the choice of local basis set will affect the result, with a choice between a less complete but more intuitive basis such as atomic orbitals and a less-intuitive but more complete basis such as maximally localised Wannier functions affecting both the accuracy and interpretation of the results. 

Now, we may rewrite the original MBD Hamiltonian with our new 6-D anisotropic oscillators instead:

Firstly, need to find the effective force-constants (the one-body term) for our oscillators

\begin{align}
    U &= \frac{1}{2}\vec{k}_{eff}(\vec{\omega})\vec{\xi}^2 \\
 \text{and also   \,}   U  &= \vec{F} \cdot \vec{\xi}\\
      &=  \mathbf{\alpha} ^{-1}(\vec{\omega}) \vec{\xi} \cdot \vec{\xi}\\
\end{align}
where $\xi$ is the displacement of the 6-D dipole oscillator, $k_{eff}$ is the effective force constant vector $\mathbf{\alpha}^{-1}$ is the inverse of  the generalised polarisability , defined as $\mathbf{\alpha} = \begin{pmatrix}
       \mathbf{\alpha}_{ee} & \mathbf{\alpha}_{em}\\
       \mathbf{\alpha}_{me} & \mathbf{\alpha}_{mm} 
    \end{pmatrix}$ in equation \ref{eqn:gen_pol}. We note that $\vec{F} \cdot \vec{\xi}$ is the energy of a dipole, $\xi$, within an electromagnetic field, $\vec{F}$.

Now, unless $\alpha^{-1}$ is diagonal (or block diagonal), we find that the electric and magnetic components of the oscillator are coupled -- even if weakly -- which is a direct result of the SOC.

\subsection{Modelling the frequency dependence of $\mathbf{\alpha}$}

So far in this paper, sparse attention has been paid to evaluating the frequency dependent nature of the polarisability. Explicit evaluation of this frequency dependence is possible \cite{Haghdani2014}, but highly computationally demanding \cite{Alliati2023}. Accordingly, we follow the original MBD method and use a model for the polarisability introduced within the original Tkatchenko-Scheffler (TS) scheme \cite{Tkatchenko2012}:

\begin{equation}
    \mathbf{\alpha}_i(i\omega) = \frac{\alpha_i(0)}{1+(\frac{i\omega}{\omega_{i}})^2}
\end{equation}
where $\alpha_i(0)$ is the static polarisability of the system, and $\omega_i$ is a characteristic frequency of  each of the oscillators, corresponding to the first (lowest frequency) resonance in the relevant component of the polarisability. This frequency corresponds to the first excitation energy of the oscillator: it qualitatively resembles a Lorentzian centred at $\omega_i$ with a value of $\alpha_i(0)$ where the function crosses the origin (i.e. the magnitude is parameterised relative to the static limit). Since the oscillator has a single excitation energy, this does not require adapting: merely evaluating. 

In a single particle picture, the first peak in the susceptibility corresponds to the HOMO-LUMO gap ($\omega_i = \frac{E_g}{\hbar}$) of the fragment. An astute reader already familiar with the drawbacks of density functional theory may point out that ``DFT underestimates the band gap'' due to the failure to correctly calculate the derivative discontinuity \cite{Perdew1982}. This in turn leads to an overestimation of the static polarisability \cite{McDowell1995} of the system since $\alpha$ is effectively being sampled at what would experimentally be a higher frequency rather than static. This drawback is, however, a symptom of an underdeveloped functional, and it may be expected that more modern functionals than the LDA  or GGAs as evaluated in ref \cite{McDowell1995} should improve on this.  For systems containing 3d transition metals or the lanthanides, a Hubbard-U correction would also be expected to significantly improve the treatment of correlation in these systems and provide a better estimate of the  static polarisability. 

\subsection{Electromagnetic Self Consistent Screening}\label{ssec:EMSCS}

The purpose of the self-consistent screening method is to account for the density response of the system to a local change -- such as rotating the dipole moment of an atomic fragment -- has on the wider system. This is not captured by a method such as the Tkachenko-Scheffler scheme \cite{Tkatchenko2009}, wherein an isolated atom is used to model the polarisability of the atom-in-molecule (or crystal). 

For our purposes, this is included within the perturbation theory evaluation of the generalised polarisability, since correct evaluation of the applied perturbation must necessarily include the full response function, which includes any off-site screenings, and any failure to do so is in fact a failure of the XC-functional of choice to correctly model the system. 

Accordingly, we can neglect the need to perform the self-consistent screening method outlined in the original many-body paper \cite{Tkatchenko2012}, at the cost of having to perform perturbative calculations instead. Whilst these calculations are more expensive, making our model prohibitively expensive for the evaluation of the charge van der Waals effect, this is a price necessary to pay in order to get reasonably accurate magnetic properties of the system.

\subsection{The Two-Body Term}\label{ssec:2body}

The starting point for evaluating the two-body interaction is the Hamiltonian $U= \vec{P}_1 \cdot \vec{F}_2$, which in the absence of SOC may be trivially separated into an electric and magnetic term. For our SOC-coupled oscillator, this is no longer the case (if an electric field can induce a magnetic displacement, i.e. $\alpha_{em}\neq 0$, then there must be an interaction with non-zero energy). 

To deal with this, we now introduce the effective electric polarisation, $\vec{p}_{eff}$, and magnetisations, $m_{eff}$ as
\begin{align}
    \vec{p}_{eff,i} &= \vec{p}_i + \alpha_{ee,i}\alpha^{-1}_{em,i} \vec{m}_i \\
    m_{eff,i} &= \vec{m}_i + \alpha_{mm,i}\alpha^{-1}_{mp,i} \vec{p}_i
\end{align}
where $\alpha^{-1}_{em,i}$ is the upper right block of the inverse of the generalised polarisability, and $\alpha^{-1}_{mp,i}$ is the lower left block of the same object.

Now, one may define the scalar potential at a point $\vec{r}$, $\Phi(\omega)$, due to an oscillating moment as
\begin{equation}
    \Phi(\omega) = \frac{{\vec{p}_{eff}} \cdot \vec{r} \omega}{4\pi\epsilon_0 c r^2} \text{sin}(\alpha)
\end{equation}
where $\alpha = \omega(t-\frac{r}{c})$ is the terms to describe the retarded potential.

The vector potential due to the oscillator at a point $\vec{r}$, may be written as $\vec{A}_{tot} = \vec{A}_{elec} +\vec{A}_{mag}$, where
\begin{equation}
    \vec{A}_{elec} = \frac{\vec{p}_{eff}\omega}{4\pi\epsilon_0 c^2 r} \text{sin}(\alpha)
\end{equation}
and
\begin{equation}
    \vec{A}_{mag} = \frac{\vec{m}_{eff} \times \vec{r} \omega}{4\pi\epsilon_0 c^3 r^2} \text{sin}(\alpha)
\end{equation}

Then, using $\vec{E} =  -\vec{\nabla} \Phi - \frac{\partial A}{\partial t}$ and $\vec{B} = \vec{\nabla}  \times \vec{A}$ we may construct the two body term as $U_{ij}= U_{elec}+U_{mag}$, where
\begin{equation}
   U_{elec} =  -\vec{p}_{eff,i} \cdot \Bigg(\vec{\nabla} \frac{{\vec{p}_{eff,j}} \cdot \vec{r} \omega}{4\pi\epsilon_0 c r^2} \text{sin}(\alpha) - \frac{\partial}{\partial t} \Bigg[\frac{\vec{p}_{eff,j}\omega}{4\pi\epsilon_0 c^2 r} \text{sin}(\alpha) + \frac{\vec{m}_{eff,j} \times \vec{r} \omega}{4\pi\epsilon_0 c^3 r^2} \text{sin}(\alpha) \Bigg]\Bigg)
\end{equation}

and 
\begin{equation}
    U_{mag} = -\vec{m}_{eff,i} \cdot \vec{\nabla} \times \Bigg[\frac{\vec{p}_{eff,j}\omega}{4\pi\epsilon_0 c^2 r} \text{sin}(\alpha) + \frac{\vec{m}_{eff,j} \times \vec{r} \omega}{4\pi\epsilon_0 c^3 r^2} \text{sin}(\alpha) \Bigg]
\end{equation}

At this point, we elect to define an operator, $T_{ij}$ such that $U_{ij} = \vec{P}_i T_{ij} \vec{P}_j$, for the sake of compactness. 

\subsection{Applied Fields}

Finally, we note that the Heisenberg model can be made to acknowledge the presence of applied fields -- which are critically important when modelling a real device or a spectroscopic interaction -- simply by adding an additional $\vec{B}_{app}\cdot\vec{S}$ term to the end of the Hamiltonian. One can likewise apply an external field to the MBD model; although in typical DFT simulations, this is likely to be set to zero.

This yields a new model Hamiltonian:

\begin{equation}
    \hat{H} = \sum_{\vec{P}_i}, \nabla^2_{\vec{P}_i} + \sum_i \vec{P}_i^T k_{\xi_i} \vec{P}_i + \sum_{i\neq j} \vec{P}_i T_{ij} \vec{P}_j + \sum_i F_\text{applied} \cdot \vec{P}_i
\end{equation}

It should also be noted that this Hamiltonian deals with the fluctuations of the spin density, not the underlying spin structure, which is a property of the static not dynamic parts of the wavefunction. This can be visualised as a permanent moment, plus a change to the moment in addition to this. This is worth noting, as this is not a consideration for a charge-density system which has no permanent point dipole moment.

\section{Comparison with the Heisenberg Model}

\subsection{Re-expression in terms of a pure Spin model}

Whilst the underlying physics of this new Hamiltonian is fully electromagnetic, for magnetic applications it would still be more useful to be able to project back into a pure spin model such as the Heisenberg Hamiltonian: both for the sake of gaining clarity on the meanings of J and K within the Heisenberg picture, and in terms of linking into existing spin dynamics models. 

 Na\"ively, this may be done by considering that commensurate excitations are generated by the same electromagnetic field, represented by pseudovector $\vec{F}$. However, $\vec{F}$ is not uniquely defined unless both $d\vec{S}$ and $d\vec{p}$ are known in advance. This in turn means that one cannot exactly reconstruct a pure spin-only model starting from the non-collinear many-body model, however various approximations may be made.

In the limit of weak spin-orbit coupling, the $\alpha_{em}$ and $\alpha_{me}$ submatrices will be approximately 0. In this case, it is reasonable to approximate there as being no hybridisation between the spin-dipole and charge-dipole waves, and one can recover both the charge-only MBD model and, by setting the kinetic energy term to 0, the Heisenberg model.

For systems with stronger spin-orbit coupling, however, it is necessary to consider the role of the off-diagonal components of the polarisability explicitly, and reconstruction of a pure spin model is impossible without drastic approximations. 

\subsection{Differences between magnons and charge-dipole waves}

\begin{figure}
    \centering
    \includegraphics[width=\columnwidth]{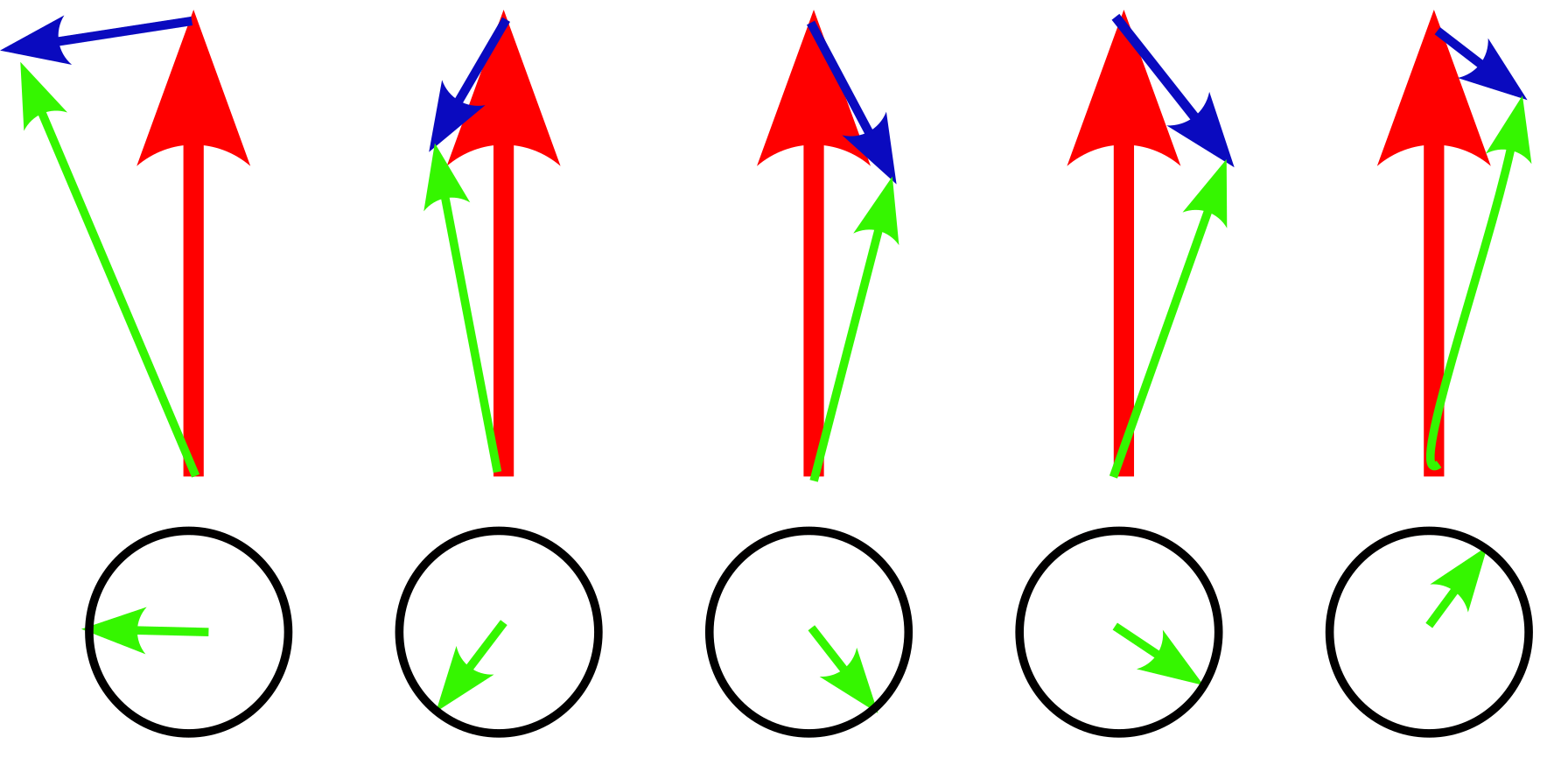}
    \caption{A sketch of magnons. The large red arrows indicate the permanent magnetic moment of the system originating in the static part of the spin density whereas the small blue arrows -- which can point in any direction -- indicate the correlated change to spin captured by the MBD-like model. This yields the thin and long green arrow which indicates the resultant magnetisation per site. The arrows in  circles below are the top-down projection of the resultant magnetisation. }
    \label{fig:magnon_mbd}
\end{figure}

\FloatBarrier
One key difference between magnons and charge-dipole waves is that whilst there is no material which has a permanent on-site  electric dipole moment (many exist where multiple atoms contribute to a dipole moment, \emph{e.g.} ionic materials and ferroelectrics), there are a wide range of materials where a permanent on-site magnetic moment exists. In this case, the observed magnon mode may be treated as a sum of the \emph{unchanging} static magnetisation plus the dynamic magnetisation, which contains all of the information about the long-distance correlations that are the magnons (so far as we can decouple these from the charge-dipole waves!). This is illustrated in figure \ref{fig:magnon_mbd}.

Now, whilst the change in magnetisation, as modelled within the MBD model, may take any value in 3-D space (noting that the model we have so far derived is only strictly applicable in the harmonic limit), the underlying magnetic ordering is likely to present a far large contribution to the resultant vector. Accordingly, the overall motion looks like the small oscillations about the static ground-state magnetic structure, as expected. A further implication of this is that atoms without a permanent magnetic moment may still support magnon modes: whilst there is no static dipole moment, the dynamical, instantaneous dipole modes may still be induced by nearby neighbours, and therefore contribute to the magnetic excitations of the system. 

\FloatBarrier
\subsection{Heisenberg's K}

In a model with spin-orbit coupling, one well-documented feature is the emergence of magnetocrystalline anisotropy as a feature of the system \cite{vanVleck1937}. Within the Heisenberg model, this is treated as an anisotropy in the one-body term, which is frequently reduced to a scalar, $k$. 

Standard methods, which apply a local constraint \cite{Lawrence2023, Liu2015} to the density should include all non-local effects, but are notably expensive, requiring a  separate self-consistent calculation at every sampled spin orientation for the system. The other popular method is tusing the magnetic force theorem, which uses a model based on an assumed functional form of the anisotropy \cite{Solovyev2021,Bruno2003} to fit a perturbation to the overall anisotropy, however this method will only reproduce the harmonic limit, leading to a poor estimation for the barrier to a spin-flip, where the anhamronic terms may be expected to be important.

This model, whilst not inexpensive, sets out a way to evaluate the harmonic limit of the anisotropy for the atoms in the system by evaluating the force-constant matrix for the oscillators in terms of the inverse of the generalised polarisability. Additionally, since the full tensor is calculated, the symmetries of the system are already included, and there is no need to assume a functional form such as uniaxial or hexagonal anisotropy; whatever the underlying symmetry of the system is is automatically included.

\subsection{Comment on Anharmonicity}

While the model presented here does not go beyond the harmonic approximation for the dipole oscillators, it is a natural extension to describe these using higher order -- anharmonic -- oscillators. In this case, the higher order anisotropies  (such as the $\text{sin}^4 \theta$ term of uniaxial anisotropy) emerge spontaneously in this model as the difference between the anharmonic terms just like the lower order anisotropy came from the  difference in the harmonic oscillators. However assessing them from an \emph{ab initio} perspective would require the calculation of the non-linear contribution to the polarisability and magnetisability, which is a beyond-DFT problem. 

Nevertheless, these anharmonic terms have been shown to lead to a simple model to predict critical temperatures \cite{Onodera1970} for classical oscillators and ferroelectric materials. It may therefore be a fruitful avenue for future study to investigate these higher order effects  to see if it is possible to determine $\text{T}_C$ and $\text{T}_N$ purely from first principles.

We note that alternative models of the van der Waals interaction such as the exchange dipole moment model (XDM) model these higher order multipoles explicitly \cite{Becke2006} (but are not typically used within a many body screening approach), which may present a natural way to parameterise the higher order anisotropies.

\subsection{Extracting Heisenberg J}

Within the context of the MBD-like model, an expression for the J-coupling between two sites can be written as

\begin{equation}
    H_{2-site} = d\vec{P}_i \mathbf{T}_{ij} d\vec{P}_j 
\end{equation}
where $\vec{P}_i$ is the generalised polarisation of site $i$, and $\mathbf{T}_{ij}$ is calculated as shown in section \ref{ssec:2body}. The parameters of this model may be calculated on any size of unit cell within a density functional theory calculation -- enabling relatively efficient calculations of the bare susceptibilities for a primitive cell (extended features such as defects would still require a supercell). 

Once the local properties have been evaluated, the rest of the evaluation, including range effects, happens inside the auxiliary model. This has the benefit that the evaluation of long-range J-couplings can be done in a supercell of the auxiliary model only, thereby facilitating extracting long-range neighbour interactions in a full tensor description at reasonable computational expense. 

Conversely, where this model is expected to be least accurate is for short-ranged interactions, due to the relatively poor handling of short-ranged interactions within the MBD scheme (we note that \emph{in principle} a range-separated screening will resolve most of these issues). These short-range couplings, however, are the easiest to evaluate using DFT methods (since only a minimal supercell is required). It should also be noted that within periodic boundary conditions, prior knowledge of the higher-order nearest neighbour couplings (e.g. second and third nearest) enables the effects of these couplings to be removed when evaluating the nearest neighbour J-couplings.

The final point of note is that the Heisenberg model is conventionally written in terms of unit vectors, whereas this model works with the magnitude of the vectors as a variable. For large amplitude oscillations (where the validity of the harmonic approximation is in doubt), the spin magnitude is known to vary by $\sim 10\%$. For small angle approximations for magnons (i.e. the harmonic limit), the spin amplitude will not vary significantly, so a Heisenberg-like J can be made from this model by direct inclusion of the relative spin magnitudes within $T_{ij}$.  

\section{Conclusion}

In this paper we have generalised the MBD model for the van der Waals interaction by substituting a full electromagnetic description for the electric-only model within the original MBD scheme. We demonstrate how the underlying physics between the Heisenberg and MBD model are linked and that they are both attempts to describe different aspects of the same physical phenomena -- the dynamic part of the electron density.

We continue to motivate how spin-orbit coupling yields (weakly coupled) spin and charge dipole waves throughout the material. Finally, we note that this enables a more direct parameterisation of Heisenberg $J$-couplings in tensor form at a lower cost than may be achieved through a na\"ive finite difference approach, as well as providing a physical motivation for the origins of anisotropy, $K$. 

\section{Acknowledgements}
The author would like to thank Peter Byrne, Phil Hasnip  for many valuable conversations and Matt Probert for supporting and encouraging the pursuit of this idea,and would also like to acknowledge UKRI (grant ref EP/V047779/1) for funding.

\bibliography{main}
\bibliographystyle{unsrt}
\end{document}